\documentclass[aps,pre,preprint,groupedaddress,showpacs]{revtex4}

\usepackage{graphicx}
\usepackage{latexsym}

\begin{document}



\title{
Stiffening of fluid membranes due to thermal undulations:
density matrix renormalization group study
}


\author{Yoshihiro Nishiyama}
\email[]{nisiyama@psun.phys.okayama-u.ac.jp}
\affiliation{Department of Physics, Faculty of Science,
Okayama University, Okayama 700-8530, Japan.}


\date{\today}

\begin{abstract}
It has been considered that the effective bending rigidity
of fluid membranes should be reduced by thermal undulations.
However, recent thorough investigation by Pinnow and Helfrich
revealed significance of
measure factors for the partition sum.
Accepting the local curvature as a statistical measure, they found
that fluid membranes are stiffened macroscopically.
In order to examine this remarkable idea, we performed extensive
{\it ab initio} simulations for a fluid membrane.
We set up transfer matrix which is diagonalized
by means of the density-matrix renormalization group.
Our method has an advantage in that it allows us to
survey various statistical measures.
As a consequence, we found that the effective bending rigidity
flows toward strong coupling under the choice of local
curvature as a statistical measure.
On the contrary,
for other measures such as normal displacement 
and tilt angle, we found clear tendency toward softening.

\end{abstract}


\maketitle


\section{\label{section1}
Introduction}

Amphiphilic molecules in water segregate spontaneously into
flexible extended surfaces called fluid (lipid) membranes 
\cite{Lipowsky91,Peliti96}.
The fluid membranes are free from both surface tension and shear modulus,
and the elasticity is governed only by bending rigidity
\cite{Canham70,Helfrich73}.
The Hamiltonian is given by the following form,
\begin{equation}
\label{Hamiltonian}
H= \int d A \left(
            \frac{\kappa}{2} J^2 + \bar{\kappa} K
           \right)   .
\end{equation}
The mean curvature $J$ is given by the summation of
two principal curvatures $J=c_1+c_2$, whereas the Gaussian curvature $K$ is
given by their product $K=c_1 c_2$.
The corresponding two moduli $\kappa$ and $\bar{\kappa}$ are 
called bending rigidity and Gaussian-curvature modulus, respectively.
The integration $\int dA$ extends over the whole membrane surface.
Hereafter, we drop the Gaussian-modulus term ($\bar{\kappa}=0$),
because this term is topologically invariant \cite{Lipowsky91,Peliti96}, and 
we restrict ourselves to a fixed topology (planar surface).

Contrary to its seemingly simple form,
the Hamiltonian (\ref{Hamiltonian}) brings about perplexing problems:
As a matter of fact, under an actual representation with differential
geometry (for instance, see Eq. (\ref{J_expression}) explained afterwards), 
it turns out that membrane's undulations are subjected to
complicated mutual scatterings.
Therefore, it is expected that the effective bending rigidity
for macroscopic length scales differs 
from the bare rigidity owing to the thermally activated undulations.
Aiming to clarify this issue, a number of renormalization-group
analyses have been reported so far
\cite{Peliti85,Forster86,Kleinert86}.
The results are summarized in the following renormalization-group
equation,
\begin{equation}
\label{renormalization_group_equation}
\kappa' = \kappa -\alpha \frac{k_B T}{8\pi}\log M         ,
\end{equation}
with renormalized bending rigidity $\kappa'$,
temperature $T$ and number of 
decimated molecules $M$.
All literatures agree that the numerical prefactor in the above equation 
is $\alpha=3$.
(More detailed account of historical overview would be
found in Ref. \cite{Helfrich98}.)
Because of $\alpha>0$, effective bending rigidity is
reduced by thermally activated undulations.
This conclusion might be convincing,
because the membrane shape itself should be deformed 
by thermal undulations.
As a matter of fact, it has been known that
the orientational correlation is lost for long distances 
\cite{deGennes82}.
It is quite natural to anticipate that membranes become floppy
for those length scales exceeding this correlation length.

Recently, however, Pinnow and Helfrich \cite{Helfrich98,Pinnow00}
obtained a remarkable conclusion $\alpha=-1(<0)$.
The key ingredient of their new argument is that they considered
the role of measure factors for the partition sum.
They insist that
the local mean curvature $J$ should be the right statistical measure
rather than other measures such as normal displacement $h(x,y)$
and local tilt angle $\theta(x,y)$.
(The variable $h(x,y)$ denotes the normal displacement
of membrane from a base (reference) plane.
We will explain it in the next section.)
After elaborated calculation of the variable replacement 
$h(x,y) \to J(x,y)$ 
and succeeding renormalization-group analysis,
the authors reach the conclusion of $\alpha=-1$.

In order to examine their remarkable scenario,
we performed first-principle simulation for a fluid membrane.
Our simulation method has an advantage in that we cover
various integration measures in a unified way.
In addition, our simulation does not rely on any perturbative
treatments.
Hence, it would be meaningful to complement the analytical
perturbative treatment.
We employed a scheme of transfer matrix, which is diagonalized
by means of the density-matrix renormalization group 
\cite{White92,White93,Peschel99,Nishino95}.
It is to be noted that recently, such elastic (bosonic) systems
come under through investigation by means of diagonalization 
after the advent of density-matrix renormalization group
\cite{Caron96,Zhang98,Nishiyama99,Nishiyama01a,Nishiyama02,Nishiyama01b}.

It has to be mentioned that the Monte-Carlo method
has been utilized successfully
in the studies of membranes and vesicles
\cite{Lipowsky91,Peliti96}.
For the Monte-Carlo method, however,
tethered (polymerized) membrane rather than fluid membrane is 
more suited \cite{Kantor86},
because membrane is implemented in computer as an assembly of molecules
and junctions bearing close resemblance to tethered membrane.
However, Gompper and Kroll succeeded in simulating 
fluid membranes by Monte-Carlo method,
allowing reconstructions of junctions during simulation
 (dynamical triangulation) \cite{Gompper00}.
They succeeded in observing the variation of topological index
with respect to the change of membrane concentration and temperature.
In fairness, it has to be mentioned that their Monte-Carlo data 
indicate softening for lipid vesicles.

The rest of this paper is organized as follows.
In the next section, we explain our transfer-matrix formalism for
a fluid membrane.
We also explicate the density-matrix renormalization group
with which we diagonalized the transfer matrix.
In Section \ref{section3},
we calculate the effective bending rigidity.
For that purpose, we introduce a scheme of coarse-graining. 
Our data clearly support membrane stiffening under the choice of
mean curvature $J(x,y)$ as a statistical measure.
For other measures, on the contrary, we observed tendency toward softening.
In the last section, we present summary and discussions.

\section{
\label{section2}
Transfer-matrix formalism and its diagonalization
through density-matrix renormalization group
}

In this section, we present our numerical simulation technique.
First, we explain our transfer-matrix formalism.
Secondly, we introduce the density-matrix renormalization group
with which we diagonalize the transfer matrix.
Demonstration of the algorithm is also presented.

\subsection{Transfer-matrix formalism}
\label{section2_1}

In order to describe the shape of membranes,
it is convenient to use the Monge gauge \cite{Chaikin95}.
In this frame, membrane deformation is described by
a normal displacement (deformation)
$h(x,y)$ from a base (reference) plane 
parameterized by Cartesian coordinates $(x,y)$.
In terms of this representation frame,
the mean curvature $J(x,y)$ is given by,
\begin{equation}
\label{J_expression}
J(x,y)=
\frac{
(\partial_x^2 h +\partial_y^2 h)(1+(\partial_x h)^2+(\partial_y h)^2)
-2\partial_x h\partial_y h\partial_x\partial_y h
-\partial_x^2 h(\partial_x h)^2
-\partial_y^2 h(\partial_y h)^2
}{
(1+(\partial_x h)^2 + (\partial_y h)^2)^{3/2}
}  ,
\end{equation}
explicitly.
Similarly,
the infinitesimal area $d A$ is given by,
\begin{equation}
d A=(1+(\partial_x h)^2 + (\partial_y h)^2)^{1/2} dx dy .
\end{equation}
Putting all together into the Hamiltonian (\ref{Hamiltonian}),
we obtain an explicit expression 
in term of the displacement field $h(x,y)$.

Now, we are led to a two-dimensional scalar-field theory $h(x,y)$.
However, the theory is afflicted by very complicated interactions.
The aim of this paper is to investigate the theory beyond
perturbative level by means of an {\it ab initio} method:
For that purpose,
we put the theory on a square lattice with lattice constant $a$;
see Fig. \ref{figure1}.
Accordingly, the field variables $\{ h_{ij} \}$ are now indexed
by integer indices $i$ and $j$.
Throughout this paper, we set the lattice constant as the unit of length $a=1$.
In other words, we are considering a square-netted membrane
which was brought into thorough discussion in Ref. \cite{Pinnow00}.
Readers may find convincing arguments why $J(x,y)$ is a physically sensible
statistical measure.
In particular, the authors think of a polymer chain, whose natural statistical
measure is the angles between adjacent links. Likewise, for a fluid membrane,
they found that
the curvatures are the right statistical measure, continuing the polymer into
a new space dimension to build up a membrane.

Our theory has the translational invariance of 
$h_{ij} \to h_{ij} + \Delta h$, and thus,
the absolute value of $h_{ij}$ should be meaningless.
Therefore, we introduce the link variable,
\begin{equation}
\label{variable_transformation}
\vec{s}= a \vec{\partial} h ,
\end{equation}
denoting the ``step'' at each link; 
see Fig. \ref{figure1}.
Obviously, we are just performing the well-known
duality transformation \cite{Villain75},
which is very successful in the study of random surfaces.
Note that now, we arrive at a dual model with step variables 
$\{ \vec{s}_{ij} \}$; see Fig. \ref{figure1}.
For this dual model, the bending-energy cost exists at each
plaquette, because it was originally a vertex possessing a curvature
spanned by adjacent links.
Hence, the statistical weight is associated at each plaquette;
\begin{equation}
\rho(s_1,s_2,s_3,s_4)  = \exp \left(  - a^2 
\sqrt{1+\left(\frac{s_1+s_4}{2a}\right)^2+
       \left(\frac{s_2+s_4}{2a}\right)^2} \frac{\kappa}{2} 
           J(s_1,s_2,s_3,s_4)^2 
                             \right)   ,
\end{equation}
with,
\begin{equation}
\label{statistical_weight_1}
J(s_1,s_2,s_3,s_4)=
-    \frac{
            (  \frac{s_4-s_1}{a^2}+\frac{s_2-s_3}{a^2} )
           \left( 1+(\frac{s_1+s_4}{2a})^2
                 +(\frac{s_2+s_4}{2a})^2 \right)  
     -(\frac{s_1+s_4}{2a})^2 \frac{s_4-s_1}{a^2}
     -(\frac{s_2+s_4}{2a})^2 \frac{s_2-s_3}{a^2}   
         }{
   \left( 1+(\frac{s_1+s_4}{2a})^2+(\frac{s_2+s_4}{2a})^2 \right)^{1.5}
         }   .
\end{equation}
See also Fig. \ref{figure1} for the notation of $\{ s_\alpha \}$.
We have set $k_B T=1$, because $k_B T$ can be absorbed into
the bending rigidity $\kappa$.
It is apparent from the construction of the dual theory that
the step variables are not fully independent.
In the notation of Fig. \ref{figure1}, there exists a restriction of 
$s_1+s_2-s_3-s_4=0$ for each open plaquette.
Therefore, we introduce the following statistical weight for each of them;
\begin{equation}
\label{statistical_weight_2}
\Delta(s_1,s_2,s_3,s_4)=\delta_{s_1+s_2-s_3-s_4,0}  .
\end{equation}

As a consequence, we reach the lattice field theory
with the statistical weights $\rho$ 
(\ref{statistical_weight_1}) and $\Delta$ (\ref{statistical_weight_2}),
 which are arranged in the checker-board pattern.
Likewise, the transfer matrix is constructed as  
a strip-like segment shown in Fig. \ref{figure1}.
The transfer matrix is diagonalized by the density-matrix 
renormalization group.
We will explain it in the next subsection.

In order to implement the above theory to computer simulation,
we must carry out yet another simplification.
Namely, we discretize the link variable as follows;
\begin{equation}
\label{step_discretization}
s_i=\delta_s (i -N_s/2-0.5)     ,
\end{equation}
with $i=1, \cdots ,N_s$.
The unit of step $\delta_s$ 
is determined selfconsistently through the simulation;
\begin{equation}
\label{selfconsistency_of_unit_of_step}
\delta_s = R \sqrt{\langle s_i^2 \rangle}          ,
\end{equation}
where $\langle \cdots \rangle$ denotes the thermal average.
We made several trials for the tuning parameters of $N_s$ and $R$;
this discretization is the most influential factor concerning
the reliability of our simulation.
We will explore
the reliability in the next subsection.

Our theory resembles the so-called solid-on-solid model,
which exhibits Kosterlitz-Thouless critical phase.
Note that in our model, there is no surface-tension term, and
there exists the bending elasticity $\kappa$ instead.
Therefore, our theory is not right at critical phase, but it is rather
driven off from it.
The renormalization-group flow is actually the central concern
of the present paper, and it is explored in the next section.

In the above, we did not pay any attention to the measure factor
for the partition sum.
As is emphasized in Introduction, the measure factor should alter 
the underlying physics even qualitatively.
As is apparent from the above formalism, 
particularly from Eq. (\ref{step_discretization}),
we accept uniform measure for the step variable;
namely, we are accepting 
the normal displacement as the statistical measure.
Following the idea advocated in Refs. \cite{Helfrich98,Pinnow00},
we will also consider the local mean curvature as for the statistical measure.
The replacement of the integration variables is absorbed into the 
redefinition of the statistical weight.
Namely, we made the replacement,
\begin{equation}
\label{statistical_weight_3}
\rho(s_1,s_2,s_3,s_4) \to \rho(s_1,s_2,s_3,s_4) 
\sqrt{
\prod_{\alpha=1}^{4}
    \left|
\frac{\partial J(s_1,s_2,s_3,s_4)}{\partial s_\alpha}
    \right|
}            .
\end{equation}
The square root is intended to take geometrical mean,
because each step variable $s_\alpha$ is sheared by an adjacent plaquette
as well.

In addition, we consider the local tilt angle as for the measure;
namely, we made the replacement,
\begin{equation}
\label{statistical_weight_4}
\rho(s_1,s_2,s_3,s_4) \to \rho(s_1,s_2,s_3,s_4)
      \prod_{\alpha=1}^4 \left|
   \cos{\rm atan}\frac{s_\alpha}{a} 
   \right|   .
\end{equation}
As a consequence, we have prepared three types of statistical weights,
$h$ (\ref{statistical_weight_1}), $J$ (\ref{statistical_weight_3})
and $\theta$ (\ref{statistical_weight_4}),
which are examined in the next section.

\subsection{Diagonalization of the transfer matrix with
density-matrix renormalization group}
\label{section2_2}

In the previous section,
we have set up the transfer matrix.
In principle, one would extract various informations, if one could 
diagonalize the transfer matrix.
However, in practice,
the matrix size exceeds the limit of computer memory size.
Such difficulty arises in common with such systems called
soft matters,
which exhibit, by nature, vast numbers of vibration modes.
Therefore, the Monte-Carlo technique has been employed 
in order to simulate soft matters.
However, after the advent of the density-matrix renormalization group
\cite{White92,White93,Peschel99},
the difficulty was removed, and now,
soft matters (elastic systems) have come under through investigations by
means of the diagonalization method; the examples are
lattice vibrations \cite{Caron96,Zhang98},
collection of oscillators as a heat bath \cite{Nishiyama99},
string meandering motions \cite{Nishiyama01a,Nishiyama02},
and lattice scalar field theory \cite{Nishiyama01b}.
In essence, the technique allows us to discard ``irrelevant states,''
and hence, the number of states are truncated so as to save the
computer-memory space.

Below, we will explain the density-matrix renormalization group.
Our algorithm is standard, and pedagogical guide would be found in 
the proceeding \cite{Peschel99} as well.
The transfer matrix is represented by a strip shown in Fig. 
\ref{figure1}.
Our goal is to diagonalize the transfer matrix with sufficient length.
We will show that this goal is achieved by the recursive application
of the density-matrix renormalization group.
The renormalization procedures are presented in Fig. \ref{figure2}, 
where two sets of renormalizations are shown.
Through each renormalization, 
a ``block'' and the adjacent site are
renormalized into a new ``block$'$;'' and similarly, 
block$' \to$ block$''$.
The crucial point is that the number of states for block
is kept within $m$.
Therefore, we can iterate the procedure until the strip 
becomes sufficiently long.
Such truncation of block states within $m$ is managed
in the following manner
\cite{White92,White93}:
One first constructs the density matrix $\rho_{B+\bullet}$ 
\cite{White92,White93,Peschel99,Nishino95}
with respect to the part of system composed of the block 
and the adjacent site.
Then,
``relevant states'' are extracted from the eigenstates 
$\{ u_\alpha \}$ of the 
density matrix with relatively larger eigenvalues $\{ w_\alpha \}$;
the density-matrix eigenvalues $\{ w_\alpha \}$ are called ``weight.'' 
Because the weight $w_\alpha$ becomes almost negligible
for large $\alpha$,
we just remain relevant $m$ states with dominant weight,
and discard the other remaining states.
In this manner, the block and the site are renormalized into a new block,
whose states are
represented by the truncated bases $\{ u_\alpha \}$ with $\alpha=1,\cdots,m$.

Let us demonstrate the reliability.
We will accept the local curvature as the statistical measure.
In Fig. \ref{figure3}, we present the distribution of weight 
$\{ w_\alpha \}$ after 40 renormalizations for bending rigidity $\kappa=0.5$. 
Simulation parameters are set to be $m=13$,
$N_s=9$ and $R=0.9$; see Eq. (\ref{step_discretization}) for details.
We see that the weight vanishes very rapidly.
In the simulation, we retain
$m=13$ weighted states, and we attain the precision of  $\sim 10^{-5}$.
In Fig. \ref{figure4}, we present the probability distribution 
of the step variable.
Because we have bounded the range of step variable
as in Eq. (\ref{step_discretization}),
we must check whether it covers the actual thermal fluctuation deviation.
The range of step variables seems to
cover the actual fluctuation deviation.
hence, our treatment of Eq. (\ref{step_discretization}) turns out to 
be justified.
In all simulations presented hereafter, we had monitored such
performance check
carefully.

\section{\label{section3}
Real-space decimation and effective bending rigidity:
application of the density-matrix renormalization group
}

In this section, we study the effective bending rigidity of a fluid membrane.
For that purpose, we introduce a scheme of real-space decimation
\cite{Swendsen82}.
Then, we apply the density-matrix
renormalization group.
All data are calculated after 40 renormalizations;
namely, the length of the transfer matrix extends to $L=80$.

To avoid confusions, we will comment a few words:
In this paper, the word ``renormalization'' is used in two different
contexts.
First, we employ the density-matrix {\it renormalization} group
as a simulation technique.
Secondly, we manage the real-space {\it renormalization} to get information of 
effective bending rigidity.
The former terminology is named after the fact that we 
{\it renormalize} irrelevant states in order to save computer-memory size.
The latter is aimed to gain the {\it renormalization}-group flow 
with length scale changed.

\subsection{Real-space decimation}
\label{section3_1}

Real-space decimation (coarse graining)
was first introduced in the study of
critical phenomena \cite{Kadanoff67}; in particular,
it was aimed to interpret the scaling hypothesis.
Later on, the idea was extended to meet more practical purposes
such as quantitative estimation of critical exponents and flow equations
\cite{Ma76}.
We follow Swendsen's version of real-space decimation,
which has been proven to be very successful \cite{Swendsen82}.

In Fig. \ref{figure5}, we have presented the 
real-space-decimation procedure.
Note that two unit cells are renormalized into one enlarged unit cell.
In other words,
two molecules are renormalized into a new molecule, and hence,
the parameter $M$ in Eq. (\ref{renormalization_group_equation}) is 
$M=2$.
Correspondingly,
from microscopic step variables $\{ s_\alpha \}$
and $\{ t_\alpha \}$, we construct the coarse-grained step variables;
\begin{eqnarray}
S_1&=&(s_1+t_1)/2   \nonumber \\
S_2&=&s_3/2+s_2          \nonumber \\
S_3&=&s_3/2+t_3         \nonumber \\
S_4&=&(s_4+t_4)/2           .
\end{eqnarray}
For this coarse-grained length scale,
the curvature is given by,
\begin{equation}
\label{coarse_grained_J}
\tilde{J}=
-    \frac{
            (  \frac{S_4-S_1}{(1.5a)^2}+\frac{S_2-S_3}{(1.5a)^2} )
           \left( 1+(\frac{S_1+S_4}{3a})^2
                 +(\frac{S_2+S_4}{3a})^2 \right)  
     -(\frac{S_1+S_4}{3a})^2 \frac{S_4-S_1}{(1.5a)^2}
     -(\frac{S_2+S_4}{3a})^2 \frac{S_2-S_3}{(1.5a)^2}   
         }{
   \left( 1+(\frac{S_1+S_4}{3a})^2+(\frac{S_2+S_4}{3a})^2 \right)^{1.5}
         }           .
\end{equation}
After that coarse-graining, one must redefine
the unit of length accordingly; namely,
\begin{eqnarray}
a &\to& a/\sqrt{2} \nonumber \\
h_{ij} &\to& h_{ij}/\sqrt{2}   .
\end{eqnarray}

The coarse-grained membrane may be described by the Hamiltonian
$H=\int dA \kappa' \tilde{J}^2/2$ with renormalized bending rigidity 
$\kappa'$.
As is well known, the transformation coefficient 
$\partial \kappa'/\partial \kappa$ contains much information
on the infrared behavior of effective coupling constant \cite{Swendsen82}.
Anticipated behaviors of $\kappa$-$\kappa'$ 
are drawn schematically 
in Fig. \ref{figure6}.
From the figure, we see that for
$\partial \kappa' /\partial \kappa>1$, the membrane is stiffened
in the infrared limit, whereas for 
$\partial \kappa' /\partial \kappa>1$, the membrane is softened.
The transformation coefficient 
$\partial \kappa' /\partial \kappa$ is given by the chain relation
\cite{Swendsen82},
\begin{equation}
\label{transformation_coefficient}
\frac{\partial \langle \tilde{J}^2 dA \rangle}{\partial \kappa}
  =
 \frac{\partial \kappa'}{\partial \kappa}
  \frac{\partial \langle \tilde{J}^2 dA \rangle}{\partial \kappa'}         .
\end{equation}
Here, $dA$ denotes the area of the membrane segment shown in Fig. \ref{figure5}.
Remaining task is to perform the above numerical derivatives.
Numerical differentiations are possible,
because our data are free from statistical errors. 
That is a great advantage of our algorithm over others such as Monte
Carlo.
We adopted the ``Richardson's deferred approach to the limit'' algorithm
in the text book \cite{NRF}.
In this algorithm, one takes an extrapolation after calculating
various finite-difference differentiations.
We monitored the relative error, and checked that the error is kept
within $10^{-3}$.

\subsection{
\label{section3_2}
Membrane stiffening in the case of local-curvature measure}

Following the idea of Refs. \cite{Helfrich98,Pinnow00},
we will accept the local curvature $J(x,y)$ as for the statistical measure.
This is achieved by adopting the statistical weight 
(\ref{statistical_weight_3}) 
in the construction of transfer matrix.
In Fig. \ref{figure7}, we plotted the transformation coefficient
$\partial \kappa' / \partial \kappa$ (\ref{transformation_coefficient})
for the bare coupling $0.17 <\kappa<4$.
In obtaining the data,
we have made a number of trials for the simulation parameters of
$m$ (number of block states),
$N_s$ and $R$ (range of step variables); see Eq. 
(\ref{step_discretization}) for details.
Such technical informations are also presented in the figure caption.
Thereby, we confirm that good convergence is achieved with respect to
the tuning parameters.

From Fig. \ref{figure7}, we see that
the inequality $\partial \kappa'/\partial \kappa>1$ holds.
Referring to the anticipated behavior depicted in Fig. \ref{figure6},
we found that the membrane is stiffened effectively for macroscopic
length scales.
As a matter of fact, for other integration measures,
the inequality $\partial \kappa'/\partial \kappa>1$ is not satisfied
as would be shown in the next subsection.
Our simulation result is the first {\it ab initio} support
of the analytical argument by Pinnow and Helfrich \cite{Helfrich98,Pinnow00}.

In the figure, for small rigidity $\kappa < 1$,
where thermal undulations should be enhanced, 
we observe notable enhancement of the transformation coefficient
$\partial \kappa'/\partial \kappa$.
Hence, it is indicated that this stiffening is driven, 
quite contrary to our naive expectation, by thermal undulations.
This rather counterintuitive result suggests that the ``hat excitation''
\cite{Helfrich98}, which is a sort of solitonic excitation, would be
created over the membrane surface:
The hat excitation is a solitonic object accompanying localized
dimple-like deformation.
In Ref. \cite{Helfrich98}, the author claimed that the hat excitations
give rise to, unlike conventional sinusoidal undulations, the membrane 
stiffening.
It would be astonishing that for such small rigidity $\kappa<0.5$, 
where the membrane
should be crumpled considerably, the concept of hat excitation is still 
applicable.
This fact may tell that the hat excitation is indeed solitonic in the sense
that the hat excitations are stable under collisions, and 
the thermal undulations
can be decomposed into individually propagating hat excitations.

It is to be noted that such pronounced enhancement of 
$\partial \kappa'/\partial \kappa$ for small $\kappa$
is not captured by the analytical one-loop renormalization-group 
treatment (\ref{renormalization_group_equation}),
because it just yields $\partial \kappa'/\partial \kappa=1$ 
for over all ${}^\forall \kappa$.
(It might be convincing, because
analytical perturbative treatment should be justified for sufficiently rigid 
membranes.)
This is obviously the advantage of our first-principle simulation.
For exceedingly small rigidity $\kappa<0.17$, however, the membrane
becomes too much crumpled, and we cannot continue reliable simulation.
(From a technical viewpoint,
the density-matrix weight exhibits almost uniform distribution, 
and we cannot set any reasonable threshold for the truncation
of states.)

On the other hand, for large-rigidity side $\kappa>1$, we see that 
the transformation coefficient approach the neutral value
$\partial \kappa' /\partial \kappa \approx 1$.
Note that 
the analytical one-loop renormalization-group analysis 
(\ref{renormalization_group_equation}) yields
$\partial \kappa' /\partial \kappa=1$.
Hence, we see that for sufficiently large $\kappa$, 
the analytical result (\ref{renormalization_group_equation}) is recovered 
asymptotically.

\subsection{\label{section3_3}
Membrane softening under the statistical measures of $h(x,y)$ and 
$\theta(x,y)$}

For other integration measures such as normal displacement $h(x,y)$
(\ref{statistical_weight_1})
and tilt angle $\theta(x,y)$ (\ref{statistical_weight_4}), 
we found behaviors quite contrastive to that shown
in the previous subsection.
In Fig. \ref{figure8}, we plotted transformation coefficient
$\partial \kappa'/ \partial \kappa$ for the normal-displacement
integration measure (\ref{statistical_weight_1}).
(Note that this integration measure has been widely used so far
in the analytical 
calculations except Refs. \cite{Helfrich98,Pinnow00}.)
From the figure, we see definitely that $\partial\kappa'/\partial\kappa<1$
 holds.
Hence, in this case, the membrane is softened effectively for macroscopic
length scales;
see the schematic behavior shown in Fig. \ref{figure6}.
In addition, it is to be noted that for small rigidity $\kappa<1$,
the transformation coefficient $\partial \kappa' / \partial \kappa$ is
suppressed, 
and it approaches the neutral value $\partial \kappa'/ \partial \kappa=1$ 
as the membrane rigidity $\kappa$ increases.
This fact tells that the softening is driven by the thermal undulations.
Note that the effect of thermal undulations appears in a way quite
opposite to
the aforementioned $J(x,y)$ statistical measure.

In the case of tilt-angle statistical measure (\ref{statistical_weight_4}),
the transformation coefficient exhibits similar behavior
$\partial \kappa'/\partial \kappa< 1$;
see Fig. \ref{figure9}.
However, 
$\partial \kappa'/\partial \kappa$ is 
much closer to the neutral value $\partial \kappa'/\partial \kappa \approx 1$, 
suggesting that
the extent of softening is smaller than that of the normal-displacement 
integration measure.
In other words, the membrane shape would stay almost scale-invariant
under the choice of tilt angle as for the statistical measure.

In the calculations presented in this subsection 
(particularly, for the $h(x,y)$ measure),
the simulations suffered from 
instabilities coming from unbounded undulations due to the
membrane softening.
More specifically, during the simulation,
membrane becomes crumpled spontaneously, 
and the membrane shape is trapped to a certain metastable configuration.
Such pathology may arise, because there are exceedingly numerous thermally
activated configurations of almost equal statistical weight.
Diagonalization of transfer matrix thus fails in searching true globally-stable
thermal equilibrium.
Moreover, the postulation of Eq. (\ref{step_discretization})
would not be fully justified, where we had assumed that the range of 
step variable is bounded.
As the simulation parameters of $m$ and $N_s$ are improved,
those instabilities are avoided to some extent.
In a sense, 
those instabilities reflect the fact that the membrane is softened indeed,
and the shape fluctuations are enhanced for long distances.
From the above experience, we are led to the speculation that 
the Monge gauge would not be very justified for the description of
fluid membranes in the case of softening, at least, beyond
perturbative level.

\section{\label{section4}
Summary and discussions}

\begin{figure}
\includegraphics{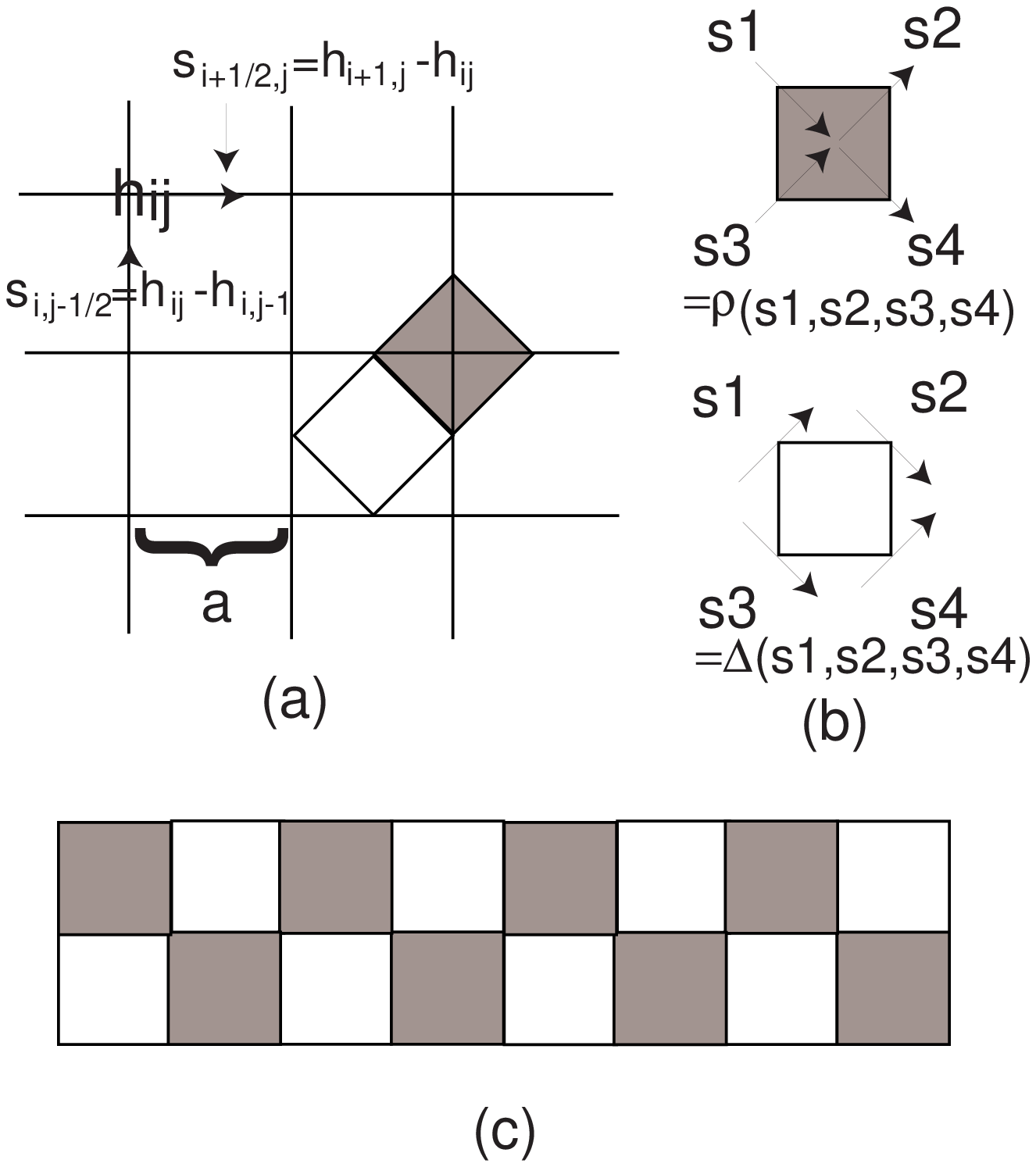}%
\caption{\label{figure1}
(a) On the square lattice,
we consider scalar field $h_{ij}$ denoting normal
displacement of a membrane with respect to a reference plane.
Step variable (gradient field) $\vec{s}=a\vec{\partial}h$ is defined
at each link.
(b) The local statistical weights $\rho$ (\ref{statistical_weight_1})
and $\Delta$ (\ref{statistical_weight_2}) are represented by shaded
and open squares, respectively.
The statistical weight $\rho$ has several variants so as to take account of
other integration measures such as curvature 
(\ref{statistical_weight_3}) and tilt angle 
(\ref{statistical_weight_4}).
(c) From those local statistical weights, we construct
a strip whose
row-to-row statistical weight yields the transfer-matrix element.
This transfer matrix is subjected to the
DMRG diagonalization as is shown in Fig.
\ref{figure2}.
}
\end{figure}

\begin{figure}
\includegraphics{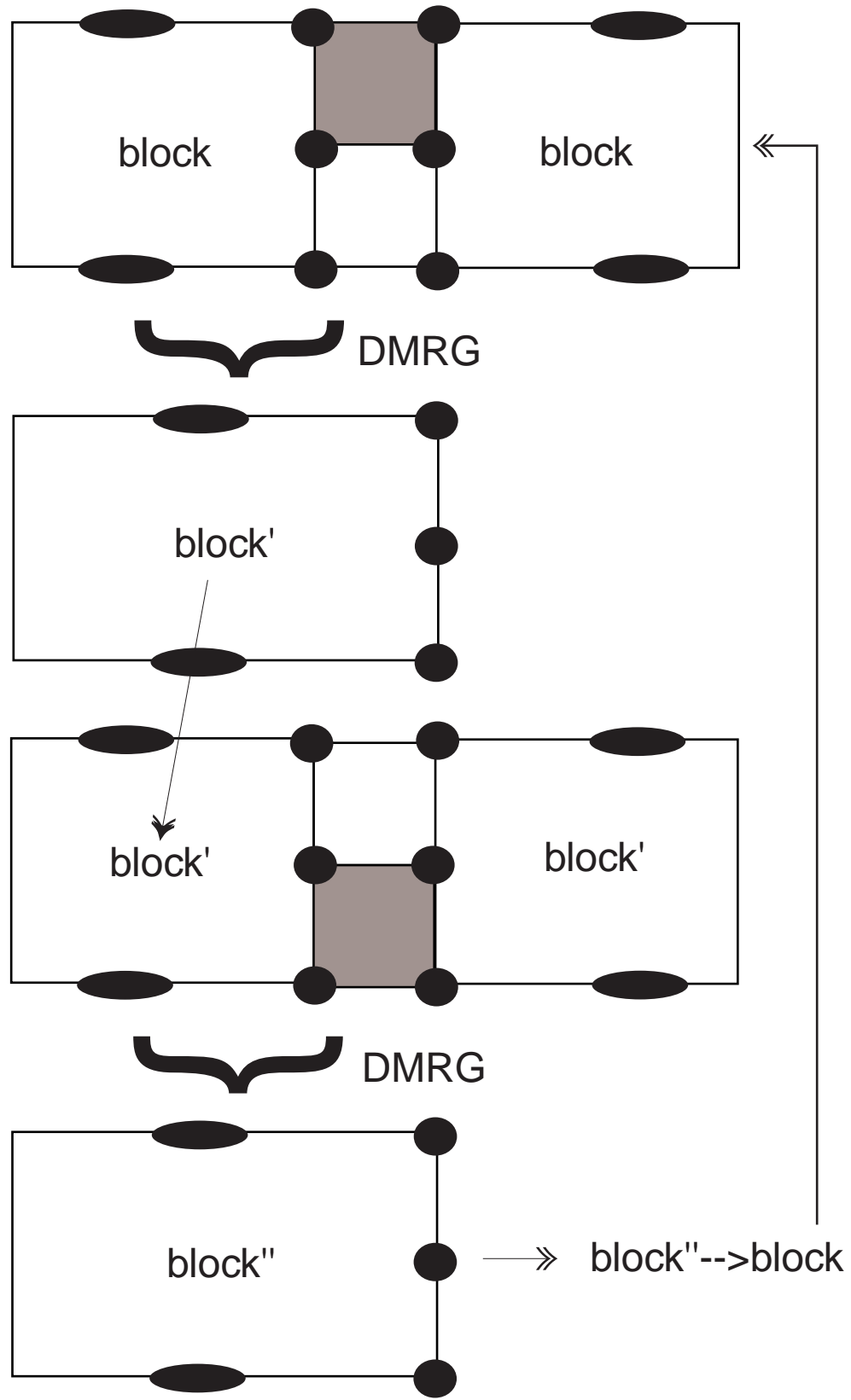}%
\caption{\label{figure2}
Schematic drawing of the density-matrix normalization group
(DMRG) procedure \cite{White92,White93,Peschel99,Nishino95}.
From the drawing, we see that through DMRG,
a ``block'' and the adjacent site is renormalized into a new ``block$'$'';
similarly, block$'$ $\to$ block$''$.
At this time, the number of block states is retained within $m$; see text.
In this manner, we can diagonalize large-scale transfer matrix
through successive application of DMRG.
Demonstration of the algorithm is presented in Figs. \ref{figure3} 
and \ref{figure4}.
}
\end{figure}

\begin{figure}
\includegraphics{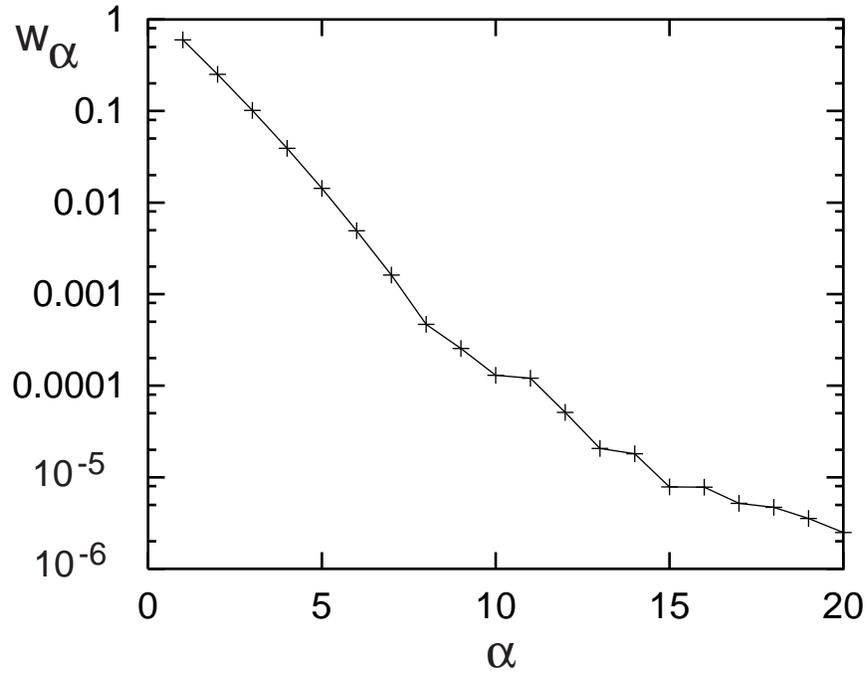}%
\caption{\label{figure3}
Distribution of the eigenvalues (weights) $\{ w_\alpha \}$ 
($\alpha$: integer index)
of the density matrix 
for bending rigidity $\kappa=0.5$.
Simulation parameters are $m=13$, $N_s=9$ and $R=0.9$; 
see Eq. (\ref{step_discretization}).
We see that $w_\alpha$ vanishes very rapidly for large $\alpha$.
In the simulation, we remained relevant states up to $m=13$, and we achieved
the precision of  $\sim 10^{-5}$.
In this manner, the number of states of ``block'' in Fig.
\ref{figure2} is truncated (renormalized).
}
\end{figure}

\begin{figure}
\includegraphics{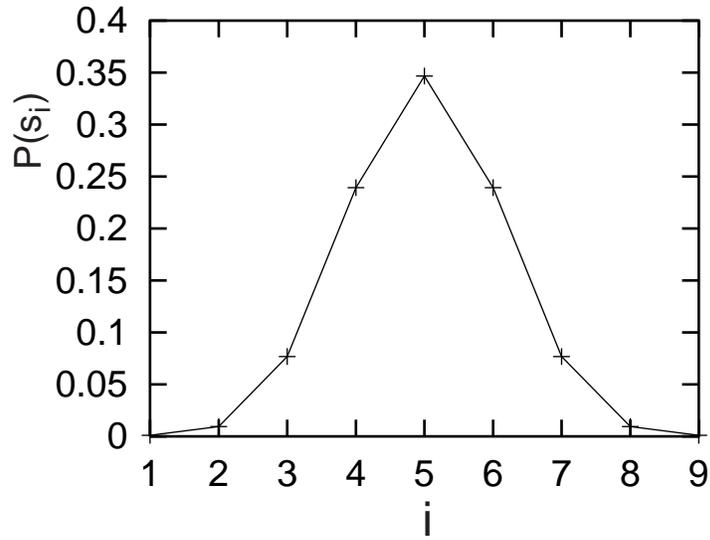}%
\caption{\label{figure4}
Probability distribution of the step variable $P(s_i)$; see Eq.
(\ref{step_discretization}), where we had set threshold for the range
of step variable like $\{ s_i \}$ ($i=1,\cdots,N_s$).
The simulation parameters are the same as those of Fig. \ref{figure3}.
The range of step variables seems to 
cover the actual fluctuation deviation.
}
\end{figure}

We have investigated the effective bending rigidity 
(\ref{renormalization_group_equation})
of
a fluid membrane
(\ref{Hamiltonian})
for macroscopic length scales.
The effective rigidity has been arousing 
renewed interest, since Pinnow and Helfrich pointed out that
the membranes would be {\it stiffened} by thermal undulations.
The key ingredient of their argument is that the local curvature
should be the right statistical measure for the partition sum.
Motivated by this remarkable scenario, we had performed
first-principle simulation with the transfer-matrix method 
and the density-matrix renormalization group; see Figs. 
\ref{figure1} and \ref{figure2}.
Our simulation scheme does not rely on any perturbative treatments,
and 
it covers various statistical measures such as
local curvature (\ref{statistical_weight_3}), normal displacement 
(\ref{statistical_weight_1}) and tilt angle 
(\ref{statistical_weight_4}).
Because analytical variable replacements among those variables require
rather tedious task even on perturbative level, 
it would be meaningful to survey various measure factors
systematically
by {\it ab initio} simulation.
Performance of the simulation scheme is demonstrated
in Figs. \ref{figure3} and \ref{figure4}.

\begin{figure}
\includegraphics{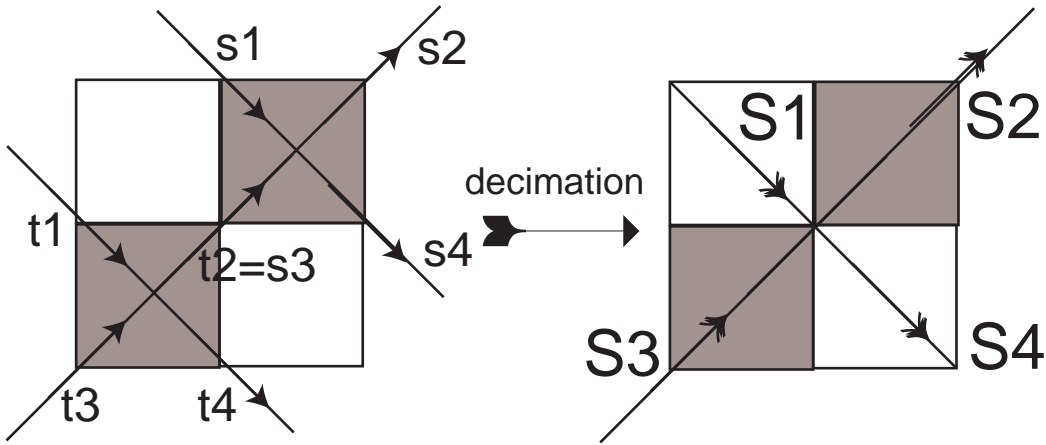}%
\caption{\label{figure5}
Real-space decimation procedure.
From the decimation, new coarse-grained curvature $\tilde{J}$ 
(\ref{coarse_grained_J}) is
constructed.
$\tilde{J}$ is used in the succeeding renormalization-group
analysis; see Fig. \ref{figure6}.
}
\end{figure}

\begin{figure}
\includegraphics{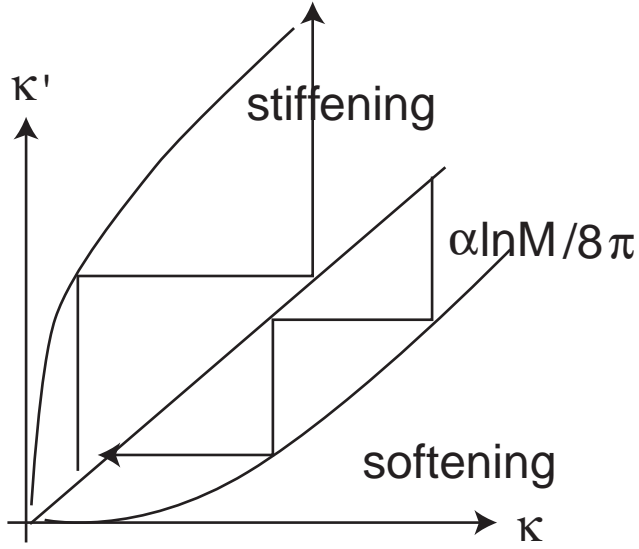}%
\caption{\label{figure6}
Schematic flow diagram of the effective bending rigidity $\kappa \to \kappa'$.
Depending on the transformation coefficient 
$\partial \kappa'/\partial \kappa>1$ ($<1$),
the membrane is stiffened (softened) in the infrared limit. 
}
\end{figure}

\begin{figure}
\includegraphics{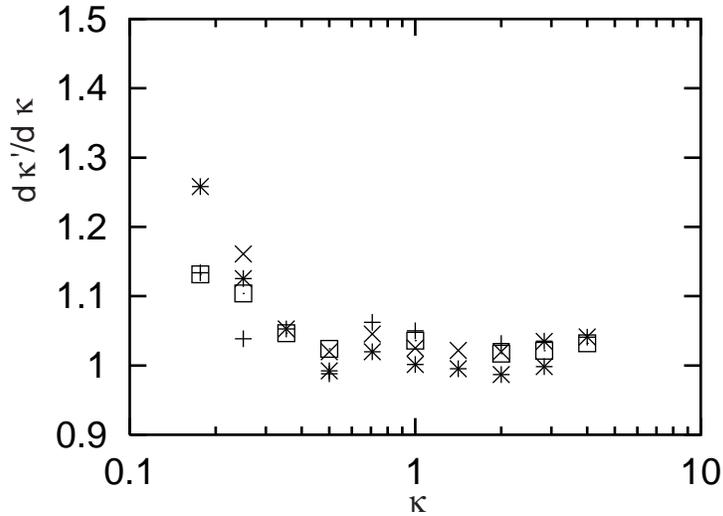}%
\caption{\label{figure7}
Transformation coefficient $\partial \kappa'/\partial \kappa$
is plotted for bare rigidity $\kappa$.
We have accepted the local curvature as for the statistical measure.
The simulation parameters for each symbol are 
($+$)      $m=12$, $N_s=7$ and $R=0.9$,
($\times$) $m=12$, $N_s=7$ and $R=1.1$ 
($\ast$) $m=11$, $N_s=8$ and $R=0.9$ and
($\Box$)   $m=15$, $N_s=6$ and $R=1.1$;
see Eq. (\ref{step_discretization}) for detail.
Referring to the anticipated behavior shown in Fig. \ref{figure6},
we see that the membrane is stiffened effectively in the infrared limit.
}
\end{figure}

\begin{figure}
\includegraphics{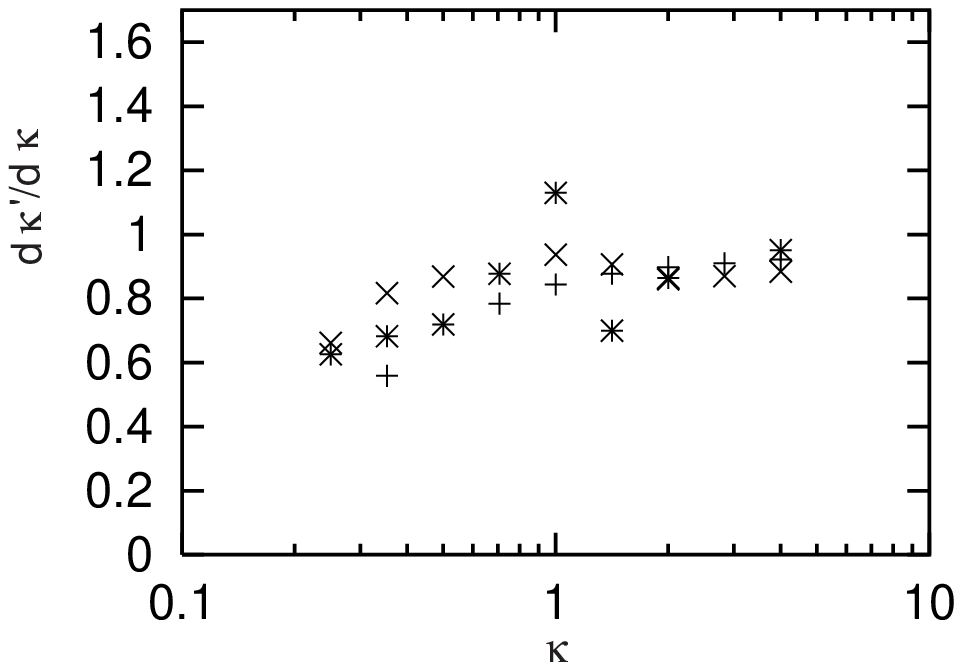}%
\caption{\label{figure8}
Transformation coefficient $\partial \kappa'/\partial \kappa$
is plotted for bare rigidity $\kappa$.
We have accepted the normal displacement as for the statistical measure.
The simulation parameters for each symbol are 
($+$)      $m=12$, $N_s=7$ and $R=0.8$,
($\times$) $m=10$, $N_s=9$ and $R=0.6$ and
($\ast$)   $m=11$, $N_s=8$ and $R=0.7$;
see Eq. (\ref{step_discretization}) for detail.
Referring to the anticipated behavior shown in Fig. \ref{figure6},
we see that the membrane is softened effectively in the infrared limit.
}
\end{figure}

\begin{figure}
\includegraphics{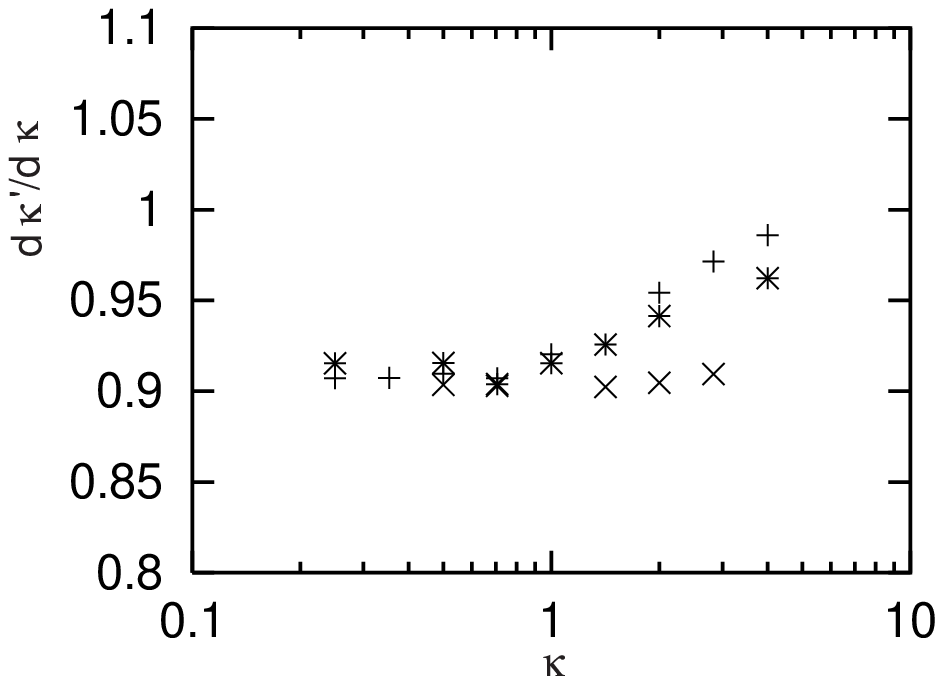}%
\caption{\label{figure9}
Transformation coefficient $\partial \kappa'/\partial \kappa$
is plotted for bare rigidity $\kappa$.
We have accepted the local tilt angle as for the statistical measure.
The simulation parameters for each symbol are 
($+$)      $m=10$, $N_s=9$ and $R=0.7$,
($\times$) $m=12$, $N_s=7$ and $R=0.7$ and
($\ast$)   $m=11$, $N_s=8$ and $R=0.7$;
see Eq. (\ref{step_discretization}) for detail.
Referring to the anticipated behavior shown in Fig. \ref{figure6},
we see that the membrane is softened effectively in the infrared limit.
}
\end{figure}

We introduced a scheme of real-space decimation as is shown in
Fig. \ref{figure5}, and correspondingly,
we defined coarse-grained curvature $\tilde{J}$ (\ref{coarse_grained_J}).
Those preparations enable us to calculate the transformation coefficient
$\partial \kappa' /\partial \kappa$ (\ref{transformation_coefficient}), 
from which we read off the 
direction of renormalization group flow;
its anticipated behaviors are drawn schematically in Fig. \ref{figure6}.
As a consequence, in respect of statistical measures,
we observed clear distinction between the mean curvature 
$\partial \kappa'/\partial \kappa>1$
and the others $\partial \kappa' /\partial \kappa<1$.
Namely, when we accept mean curvature as a statistical measure,
the effective rigidity is, in fact, renormalized toward strong
coupling; see Fig. \ref{figure7}.
That is,
membrane stiffening takes place.
On the contrary, both
normal displacement and tilt angle appear to lead to membrane softening; 
see Figs. \ref{figure8} and \ref{figure9}.
Our simulation results are the first {\it ab initio} support of the
aforementioned analytical argument \cite{Helfrich98}.

The membrane stiffening,
apart from its mere curiousness, is quite favorable for the consistency
of our numerical simulation:
In our simulation, in the first place,
we had restricted the range of step variables as in
Eq. (\ref{transformation_coefficient}). 
Apparently, this postulation is in favor of membrane stiffening.
In addition, we had discarded irrelevant states
through the density-matrix renormalization group in order to keep
the number of states tractable in computers.
Again, this truncation of states is validated consequently
after the onset
of membrane stiffening.
Furthermore, we believe that the very starting point of our theory,
namely, the Monge gauge, is in favor of membrane stiffening,
because it is assumed that there exists a reference plane 
from which all undulations are excited.
Membrane stiffening may serve solid grounds for 
the Monge gauge.

For those reasons, we could perform well-controlled simulation 
in the case of membrane stiffening.
On the contrary, as for the case of membrane softening,
our simulation faces a number of conflicts coming from 
unbounded undulations as the strip length of transfer matrix is enlarged;
see Section \ref{section3_3} for details.
Because such pronounced undulations deprive 
the grounds of the reference plane, it would be rather suspicious
to adopt the Monge gauge at least beyond perturbative level.
By the way, we are fairly confident of the efficiency of our simulation
in the case of membrane stiffening, and we believe that
the membrane stiffening may serve a new promising research field 
for the application of the density-matrix renormalization group.

The following problems remain open:
First, In the article \cite{Helfrich98}, the author raised an intriguing 
picture of deformation mode called ``hat excitation.''
This is a sort of soliton which is in contrast to the naive sinusoidal
undulation.
With the hat-excitation picture, the author gave valuable physical 
insights.
First-principle
examination of the reality of this hat excitation would be desirable.
Secondly, irrespective of the membrane stiffening, the correlation length
is known to be finite, and the orientational correlation is lost for
long distances \cite{deGennes82}.
In Ref. \cite{Helfrich98}, it is speculated that those two modes,
namely, orientation and curvature, would be decoupled.
A deeper understanding of the above points would establish the 
justification 
that the local curvature is the right statistical measure.

\begin{acknowledgments}
This work is supported by Grant-in-Aid for
Young Scientists
(No. 13740240) from Monbusho, Japan.
\end{acknowledgments}


\end{document}